\documentclass[prd,aps,reprint,twocolumn,superscriptaddress,tightenlines,nofootinbib,eqsecnum,preprintnumbers,longbibliography]{revtex4-2}
\usepackage[usenames,dvipsnames]{xcolor}
\usepackage{epsfig,amsfonts,mathrsfs,amsmath,amssymb,graphicx,color,slashed,multirow}
\usepackage{amsmath,latexsym,amssymb,graphicx,slashed,color,enumerate,url,etoolbox,cancel,gensymb,physics}

\usepackage[usenames,dvipsnames]{xcolor}
\definecolor{nicered}{rgb}{0.7,0.1,0.1}
\definecolor{nicegreen}{rgb}{0.1,0.5,0.1}
\definecolor{niceblue}{rgb}{0.1,0.2,0.6}

\usepackage[colorlinks=true,urlcolor=nicered
,anchorcolor=nicered,citecolor=nicegreen,filecolor=nicered,linkcolor=nicered,menucolor=nicered
]{hyperref}

\usepackage{bbm}
\usepackage{physics}
\usepackage{dsfont}
\usepackage{cancel}
\usepackage{float}
\usepackage[caption=false]{subfig}

\newcommand{\mhpl}{m_{H^+}}

\begin{document}

\def\TRIUMF{TRIUMF, 4004 Wesbrook Mall, Vancouver, BC V6T 2A3, Canada}
\def\SFU{Department of Physics, Simon Fraser University, Burnaby, BC V5A 1S6, Canada}

\title{The Sign of Gauge-Higgs Couplings at Future Lepton Colliders}

\author{Carlos Henrique de Lima}
\email{cdelima@triumf.ca}
\affiliation{\TRIUMF}
\author{Douglas Tuckler}
\email{dtuckler@triumf.ca}
\affiliation{\TRIUMF}
\affiliation{\SFU}

\date{\today}

\begin{abstract}
This work investigates the capability of future lepton colliders to determine the sign of the gauge-Higgs coupling through the vector boson fusion (VBF) $Zh$ process. This channel offers a model-independent way to probe the sign of the gauge-Higgs coupling. Its sensitivity to interference effects and universal coupling with new physics makes it particularly effective. We show that a high-energy lepton collider such as CLIC can fully determine the sign of the gauge-Higgs coupling in a model-independent way and with high confidence.
\end{abstract}

\maketitle

\section{Introduction}

\begin{figure*}
\centering
\subfloat[\label{fig:Zhcross}]{
\includegraphics[width=0.485\linewidth]{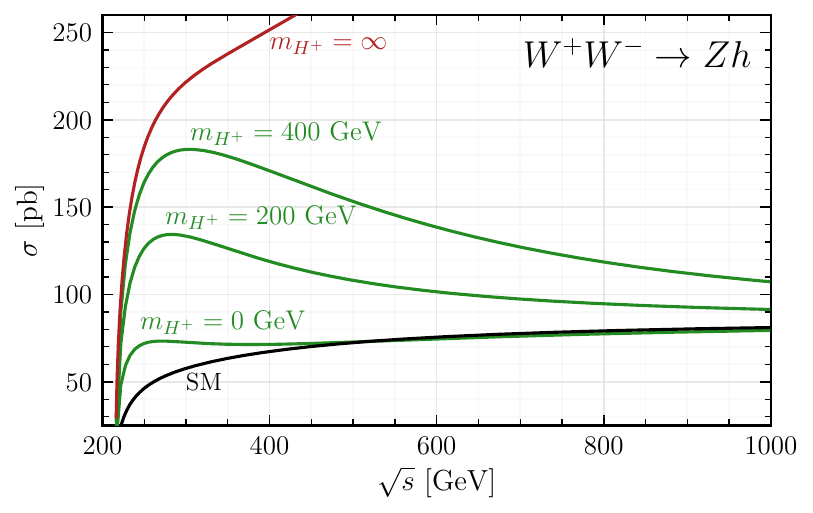}}
\subfloat[\label{fig:crossDEC}]{
\includegraphics[width=0.485\linewidth]{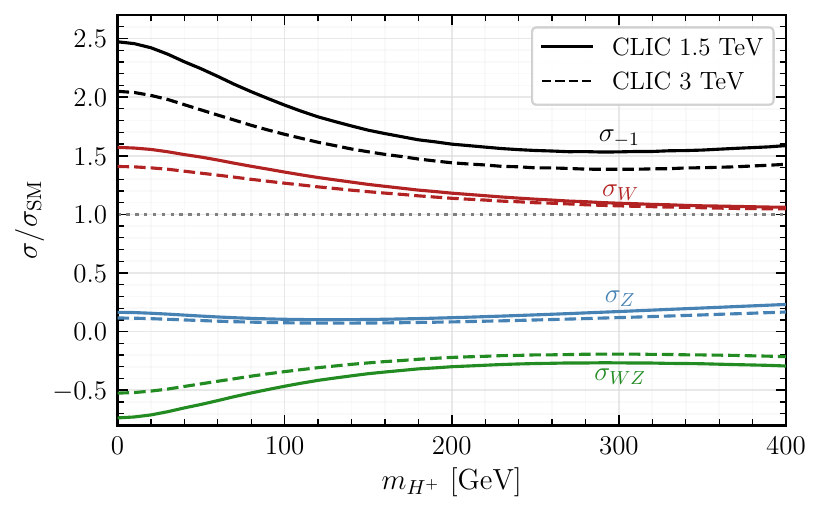}}
\caption{(a): Cross section for $W^{+}W^{-}\rightarrow Zh$ as a function of the center-of-mass energy $\sqrt{s}$ with only one charged Higgs state. The green curves correspond to different charged Higgs masses, while the red curve is the heavy mass limit $m_{H^+} \to \infty$. The black curve is the SM prediction, i.e., with $\kappa_W = \kappa_Z = 1$. The charged Higgs couplings are fixed by unitarization. (b): Cross section decomposition described in Eq.~\eqref{eq:dec} as a function of the charged Higgs mass $m_{H^+}$, normalized to the SM cross section $\sigma_{\text{SM}}$. The solid lines are for $\sqrt{s} =$ 1.5 TeV and the dashed lines for $\sqrt{s} =$ 3 TeV, both with CLIC polarization ($P(e^{-})=-80\%$). Note, $\sigma_{-1}$ denotes the total cross section with $\kappa_W = -\kappa_Z = 1$.}
\label{fig:figure1}
 \end{figure*}

Since the discovery of the Higgs boson at the LHC~\cite{higgsDISCATLAS,higgsDISCCMS}, diverse measurements indicate that the properties of the Higgs boson are consistent with Standard Model (SM) predictions (see \cite{ATLAS:2022vkf,CMS:2022dwd} for recent reviews). This suggests that new physics (NP) effects in the Higgs sector must be small. However, an intriguing possibility is that NP effects are not small, but instead hidden from experiments. In particular, the determination of Higgs couplings to SM particles from rate measurements is only sensitive to the magnitude of the couplings. Thus, NP effects that change the sign of Higgs couplings to SM particles are not easily differentiated from the SM, and new strategies must be used to measure the sign of Higgs couplings. Beyond understanding Higgs properties, determining the sign of the Higgs couplings has profound implications for our understanding of electroweak symmetry breaking.

The relative sign of the Higgs couplings is only probed through interference effects in, for example, Higgs decays to four leptons~\cite{Chen:2016ofc}, $W^+W^- h$ production~\cite{Chiang:2018fqf}, $Vh$ production via vector boson fusion (VBF)~\cite{Stolarski:2020qim}, and the combination of $Zh$ and $th$ production~\cite{Xie:2021xtl}. The challenge with interference channels is that their analysis is sensitive to potential EW scale physics, which is unaccounted for in the standard coupling modifier framework \cite{LHCHiggsXSWG:2013rie}. These wrong sign scenarios significantly change the Higgs couplings, implying that at least one new state must be below about 620 GeV~\cite{Das:2024xre} and thus could significantly affect the interference measurements.

The sensitivity of the interference analysis to NP was pointed out in~\cite{deLima:2024uwc} in the context of the gauge-Higgs coupling sign using the VBF $Wh$ process. The gauge-Higgs coupling is extra sensitive to NP because of its crucial role in longitudinal gauge boson scattering unitarity. In~\cite{deLima:2024uwc}, it was shown that any generic ultraviolet (UV) completion (as constructed in~\cite{deLima:2021llm}) that can generate a negative gauge-Higgs coupling would inevitably have new charged states contributing to the VBF $Wh$ process. A recast of the ATLAS analysis~\cite{ATLAS:2024vxc} done in~\cite{deLima:2024uwc} shows that a previously excluded high confidence hypothesis is now allowed if there is at least one charged Higgs with mass below $370~\text{GeV}$. A similar bound is expected from the CMS data~\cite{CMS:2023sdc}.

Interestingly, the analysis cannot reach lower masses because, in this regime, the cross-section and kinematics are too similar to the SM. This means that under current experimental data, the negative coupling hypothesis is not excluded but requires at least one charged Higgs state below $370~\text{GeV}$. The search for such a scalar is heavily model-dependent, and thus, it is always possible to engineer a UV completion that is not excluded. 

In order to thoroughly probe the negative gauge-Higgs coupling hypothesis, it is necessary to have a \emph{model-independent} reach for the charged Higgs state below $370~\text{GeV}$. This is where the VBF $Zh$ process enters. This process is more sensitive to low masses of the charged Higgs and thus can be used to thoroughly probe the negative gauge-Higgs coupling scenario. At the Large Hadron Collider (LHC), the cross section for this process is 20 times smaller than VBF $Wh$, making it challenging even at HL-LHC~\cite{ATLAS:2019mfr}. 

In this paper, we determine the sensitivity of future high-energy lepton colliders to the sign of the gauge-Higgs couplings using $Zh$ production via VBF

\begin{align}\label{eq:proc}
\ell^{+} \ell^{-} \rightarrow \nu_{\ell} \bar{\nu}_{\ell} Z h.
\end{align}

We focus on the Compact Linear Collider (CLIC) at center-of-mass energies $\sqrt{s} = $1.5 TeV and 3 TeV \cite{Aicheler:2018arh,Zarnecki:2020ics}, with polarized $e^-$ beam ($P(e^-) = -80\%$). Even though we focus on CLIC, any other high-energy future collider can optimally probe the sign of the gauge-Higgs coupling, provided enough luminosity and energies higher than TeV. The necessity of higher energy usually implies later stages of the different collider proposals~\cite{Bhat:2022hdi,Narain:2022qud} like the International Linear Collider (ILC)~\cite{Bambade:2019fyw,Zarnecki:2020ics,Behnke:2013xla,Adolphsen:2013jya,Adolphsen:2013kya}, the Cool Copper Collider (C$^3$)~\cite{Vernieri:2022fae,Breidenbach:2023nxd}, and future Muon Collider~\cite{Andreetto:2024rra}. We also point out that polarization is not critical for this channel, where the same result drawn from $P(e^-) = -80\%$ can be obtained with double luminosity with an unpolarized beam for the energies studied.

The remainder of this paper is organized as follows. In Sec.~\ref{sec:MIan}, we review the model independence of the $VV \rightarrow Vh$ process and the parametrization in terms of the $\kappa$ framework with one additional charged Higgs state. In Sec.~\ref{sec:VBF}, we analyze the effect of charged Higgs states on the VBF $Zh$ cross-section. In Sec.~\ref{sec:pheno}, we perform the phenomenology study and show the possibility of probing the negative gauge-Higgs hypothesis in the near future at a high-energy lepton collider such as CLIC. We conclude in Sec~\ref{sec:conc}.

\section{Model independent analysis of VBF $Zh$}\label{sec:MIan}
\begin{figure*}[t]
\centering
  \resizebox{0.475\linewidth}{!}{ \includegraphics{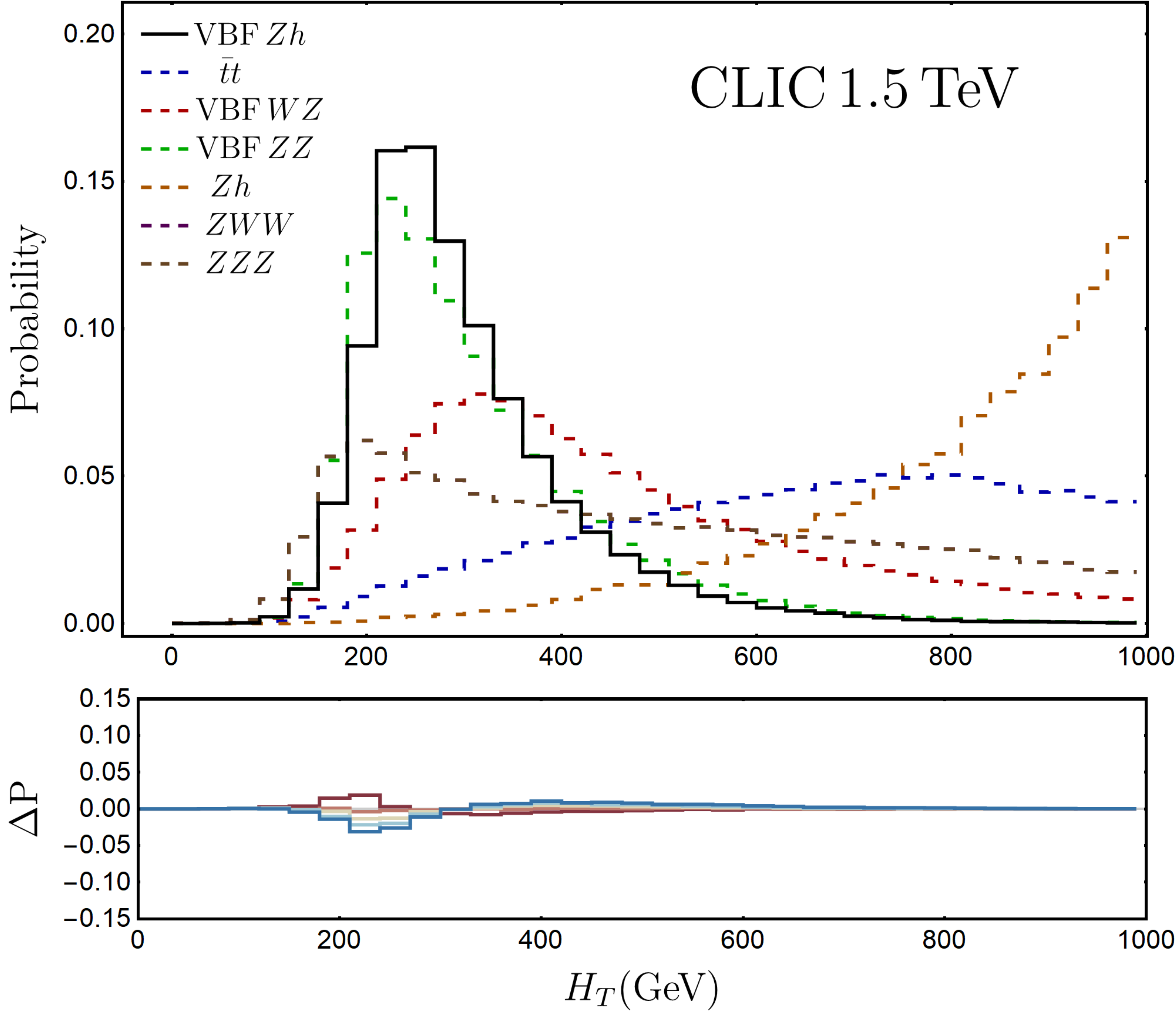}} 
  \resizebox{0.475\linewidth}{!}{ \includegraphics{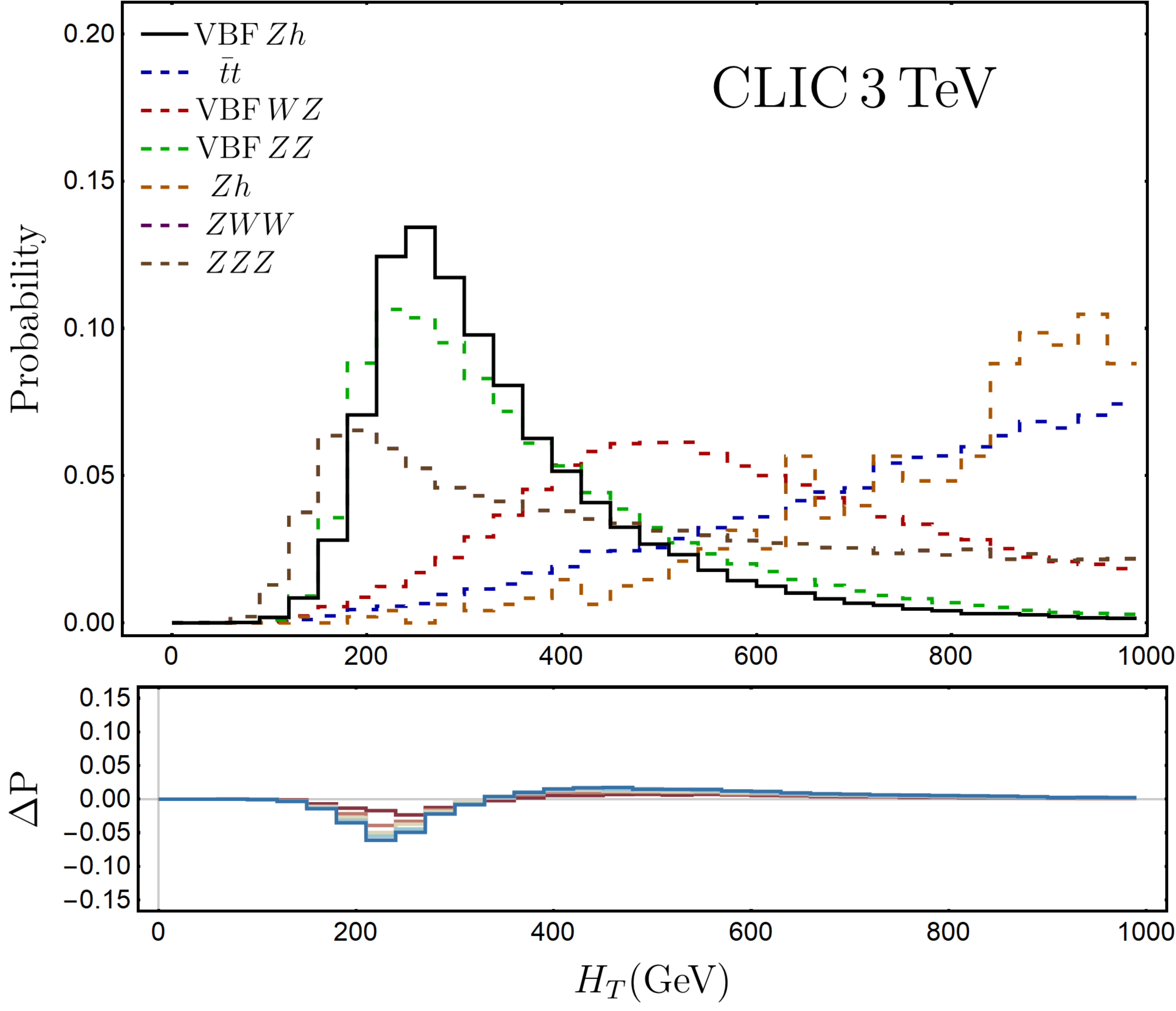}} 
\caption{Normalized $H_{T}$ distribution of the VBF $Zh$ signal process with $\lambda_{WZ}=1$ (solid black) and various backgrounds (dashed) for 1.5 TeV (left) and 3 TeV (right) CLIC with polarization ($P(e^{-})=-80\%$). In the bottom, we have the difference between the distribution for $m_{H^+}= 0,100,200,300,400$ GeV (ordered from red to blue) at $\lambda_{WZ}= -1$ and the SM distribution (solid black line in the upper plots) showing the small change in kinematic distributions for different $m_{H^+}$.  }
\label{fig:kins}
\end{figure*}

We begin by reviewing how charged Higgs states enter the $ VV \rightarrow Vh$ scattering in the context of extended Higgs sectors, where $V$ is either a $W$ or $Z$ boson. The modification of the gauge-Higgs couplings is described model independently in the $\kappa$-framework as

\begin{equation}
\begin{aligned}
\mathcal{L}_{\kappa}=h\left(\kappa_W   g m_W  W_\mu^{+} W^{-\mu} + \kappa_Z  g \frac{m_Z^2}{2 m_W}   Z_\mu  Z^\mu\right)
\end{aligned}
\end{equation}

\noindent where $m_{W,Z}$ are the $W$ and $Z$ boson masses, $g$ is the $SU(2)_L$ coupling constant, and $\kappa_{W,Z}$ are the gauge-Higgs coupling modifiers defined as $\kappa_i ^2= \sigma_i/\sigma_i^\text{SM}$ \cite{LHCHiggsXSWG:2013rie}. We additionally define the ratio of the coupling modifiers $\lambda_{WZ} \equiv \kappa_W/\kappa_Z$.

In the SM, $\lambda_{WZ} = 1$, but this relation need not hold in extensions of the SM, for example, in UV theories with custodial violation. The construction of these scenarios involves scalars in higher $SU(2)_{L}$ representations, which introduce additional charged states after electroweak symmetry breaking. For the $ VV \rightarrow Vh$, only a singly charged state contributes, and we can write the most general custodial violating interaction, which is relevant for the process as
\begin{equation}\label{Eq:NPEQ}
\begin{aligned}
\mathcal{L}_{H^{+}} = \sum_{i} g  m_{Z}  g^{i}_{WZ} H_{i}^{+}W^{-}_{\mu}Z^{\mu} - i  g  g^{i}_{Wh}  h  \partial_{\mu}H^{+}_{i}  W^{-  \mu} \, .
\end{aligned}
\end{equation}

Independent of any particular UV completion, unitarity of the scattering cross section requires the presence of charged Higgs states. For example, in the $W^+ W^- \rightarrow Z h$, the amplitude for longitudinally polarized gauge bosons has the following high energy limit
\begin{align}
    \mathcal{M}_{Zh}^{LLL} &= \frac{g^2\cos\theta}{4m_{W}^{2}} \left(\sum_{i}g^{i}_{WZ}g^{i}_{Wh} +\kappa_{W} -\kappa_{Z} \right) s + \mathcal{O}(s^0),
     \label{eq:mLLL}
\end{align}
where the first term is the $t$-channel contributions from the charged Higgs bosons\footnote{ We assume that CP is a good symmetry of the scalar sector, which forbids a neutral scalar exchange in this process. CP is expected to be only an approximate symmetry~\cite{deLima:2024hnk,deLima:2024lfc} from the custodial breaking potential, but the coefficients of CP violating terms will be heavily suppressed and can be phenomenologically ignored.}. To prevent perturbative unitarity violation, any consistent extended Higgs sector must satisfy the sum rule
\begin{align}\label{eq:unit}
\sum_{i}g^{i}_{WZ}g^{i}_{Wh} =\kappa_{Z} -\kappa_{W}  . 
\end{align}

This model independence allows us to test the negative gauge-Higgs hypothesis using the $\kappa-$framework with  $i$ charged Higgs scalars satisfying Eq.~\eqref{eq:unit}. These new states control the high-energy behavior of the scattering amplitude but generate significant contributions at low energies. This sum rule shows that the high energy behavior is controlled in the $s\rightarrow \infty$ limit. However, depending on the mass of the charged state, unitarity is already violated at much lower, collider-accessible energies. Each state $i$ contributes with $g^{i}_{WZ}g^{i}_{Wh}$ for the amplitude only when its kinematically accessible. We can see this behavior in Fig.~\ref{fig:Zhcross}, where the heavier the charged state becomes, the later the high energy behavior is tamed. 

Generally, a UV completion can have more than one charged Higgs state. Each state can only contribute to the unitarization of the amplitude once the energy probed is higher its mass. If no state is below $370~\text{GeV}$ from this picture, we recover the ATLAS and CMS exclusion~\cite{deLima:2024uwc}. Therefore, we can look for only one state in this range, and if this is experimentally excluded, then the scenarios with multiple charged states are also automatically excluded. Excluding the one-state scenario means excluding the negative gauge-Higgs coupling hypothesis. We show this explicitly for the two-state scenario in Appendix~\ref{sec:ap}.

This work focuses on the $Zh$ final state as this channel is more sensitive to light charged Higgs states. We look only for states below $370~\text{GeV}$ as both ATLAS and CMS have excluded model-independently the heavier charged states using VBF $Wh$. In the next section, we show that a future lepton collider has sensitivity in this parameter region and that the sign of the gauge-Higgs coupling can be model independently determined.

\section{VBF $Zh$ production}\label{sec:VBF}
We consider $Zh$ production through vector boson fusion (VBF) at a future lepton collider:
\begin{align}
e^+ e^- \rightarrow \nu_{e} \bar{\nu}_{e} Z h \, .
\end{align}
The analysis is similar to the one performed in~\cite{Stolarski:2020qim}, with the main difference that there is an additional contribution from charged Higgs states when $\kappa_{W} \neq \kappa_{Z}$. The SM cross section is sensitive to the initial polarization, enhancing the cross section by a factor of two when the electron is negatively polarized. 

We explore VBF $Zh$ production at CLIC with a polarization of $P(e^{-})=-80\%$ and a total luminosity of 1 ab$^{-1}$, sufficient for a high confidence bound on the negative coupling regime. In addition, we consider CLIC running at $\sqrt{s} =$ 1.5 and 3 TeV. We have explored the possibility of ILC at 500 GeV with 1.5 ab$^{-1}$ and verified that it does not have good reach, mostly because of the lower center of mass energy. The backgrounds at lower energies are dominated by $t\bar{t}$ and the signal is sensitive to the energy from the VBF topology.  We expect that any lepton collider with center-of-mass energy larger than 1 TeV can effectively probe this channel with significant statistics.

\begin{figure*}[tbh]
\centering
\includegraphics[width=0.328\linewidth]{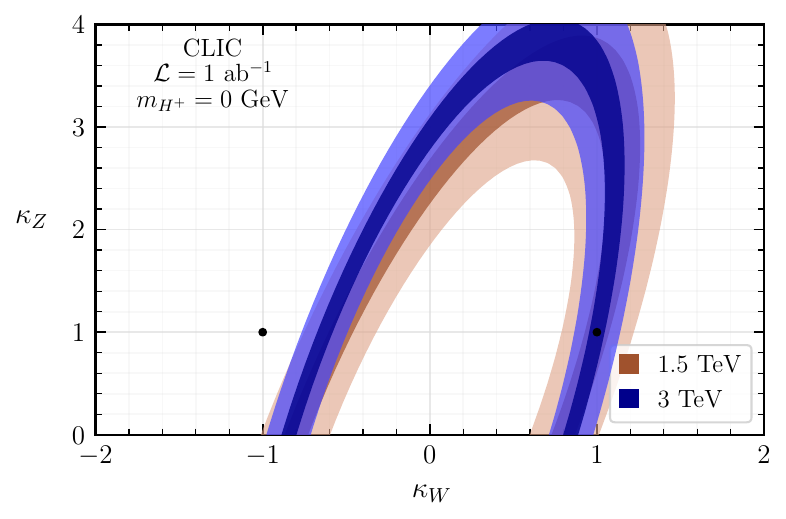} 
\includegraphics[width=0.328\linewidth]{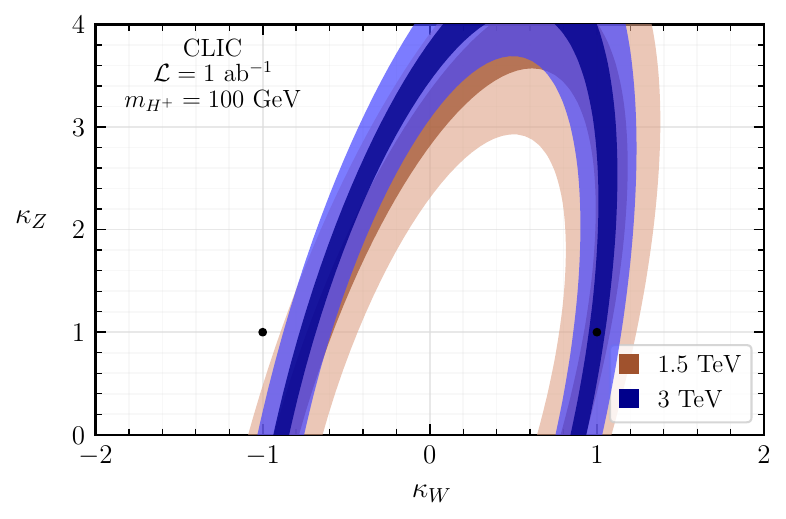}
\includegraphics[width=0.328\linewidth]{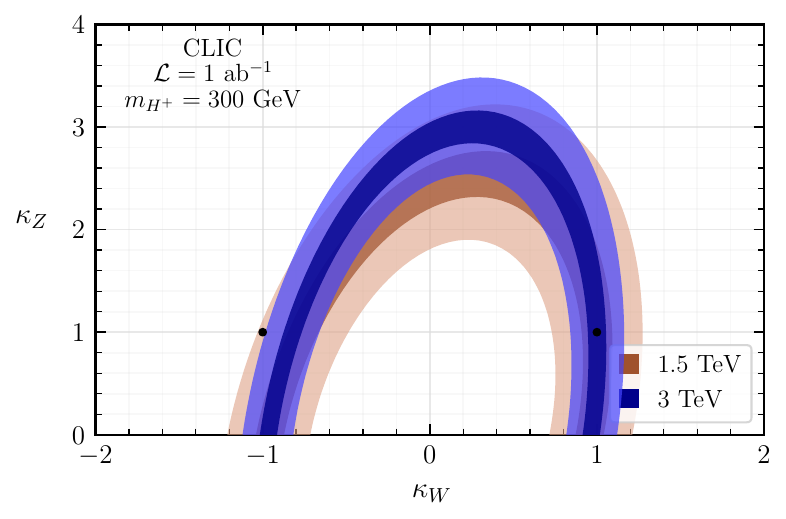}
\caption{Sensitivity of CLIC with $\mathcal{L}=1 \text{ab}^{-1}$ at $\sqrt{s} = 1.5$ TeV (brown) and $\sqrt{s} = 3$ TeV (blue), for different charged Higgs masses (left to right) $m_{H^{+}} = (0, 100, 300)$ GeV. The darker (lighter) bands represent the sensitivity of $2\sigma$ ($5 \sigma$).}
\label{fig:moneyplot}
\end{figure*}

We parametrize the VBF cross section similar to~\cite{Stolarski:2020qim}, but we include the charged Higgs mass dependence
\begin{equation}
\begin{aligned} \label{eq:dec}
\sigma(\kappa_W,\kappa_Z,m_{H^{+}}) = \kappa_{W}^{2} &\sigma_{W}(m_{H^{+}}) + \kappa_{Z}^{2} \sigma_{Z}(m_{H^{+}}) \\
&+ \kappa_{W}\kappa_{Z} \sigma_{WZ}(m_{H^{+}})  \,,
\end{aligned}
\end{equation}
where $\sigma_{W}$ and $\sigma_Z$ denote contributions to the cross section from $hWW$ and $hZZ$ couplings, respectively, while $\sigma_{WZ}$ is the interference term.

The cross section for $W^+W^- \to Zh$ as a function of $\sqrt{s}$ is shown in Fig.~\ref{fig:Zhcross} for different charged Higgs masses and $\lambda_{WZ} =-1$ (green and red curves). The charged Higgs couplings are fixed by Eq.~\eqref{eq:unit}. When $m_{H^+} = \infty$ (red curve), the cross section grows with energy as expected. For a finite charged Higgs mass (green curves), the cross section is finite and enhanced compared to the SM prediction with $\lambda_{WZ} = 1$ (black curve). This enhancement has a non-trivial effect on the event rate at colliders. 

In Fig.~\ref{fig:crossDEC}, the decomposition of the VBF $Zh$ cross section as a function of the charged Higgs mass is shown at CLIC for  $\sqrt{s} = 1.5$ and 3 TeV. The cross sections are normalized to the SM cross section $\sigma_\text{SM}$. For $\lambda_{WZ} = 1$ we recover the SM cross section and $\sigma/\sigma_\text{SM} = 1$. However, the total cross section is always higher than the SM for $\lambda_{WZ} = -1$. This is depicted by the black curves labeled $\sigma_{-1}$. In addition, there is a value of $\mhpl$ for which the negative-coupling cross section is a minimum. As we implement cuts in kinematic variables that are independent of $\mhpl$, excluding this point translates into an exclusion of all charged Higgs masses. 

Note, the ratio $\sigma_{-1}/\sigma_\text{SM}$ decreases with the increasing center-of-mass energy. This trend continues as the energy is increased to 10 TeV, where the smallest ratio is $\sim 1.25$. This shows that the process requires a minimal energy to be probed, but after that it is mostly luminosity dependent since it becomes harder to distinguish the two hypotheses at high energies.

 In the following section, we perform a detailed phenomenological study of this channel, including backgrounds and realistic experimental cuts.

\section{Phenomenology Study}\label{sec:pheno}

A measurement of the VBF $Zh$ cross section can be interpreted as a measurement of $\lambda_{WZ} \equiv \kappa_W/\kappa_Z$ and $m_{H^+}$. Since this process has increased sensitivity for the small mass regime, we can probe all $m_{H^+}$ to ultimately determine the sign of the gauge-Higgs coupling. We study this cross section measurement at CLIC with an integrated luminosity of 1 ab$^{-1}$ and center-of-mass energies at $\sqrt{s} =$ 1.5 and 3 TeV. We construct the fiducial acceptance and include a total efficiency factor to account for the detector and tagging efficiency. While we consider a linear $e^+ e^-$ collider for this study, the analysis is similar for a muon collider.

 The signal process we consider is
\begin{align}
e^+ e^- \rightarrow \nu_{e} \bar{\nu}_{e} Z h \,,
\end{align}
with the final state containing two same-flavor leptons from the $Z$ decay, two $b$-jets from the Higgs decay, and missing energy. The dominant backgrounds are from:

\begin{align}
e^{+} e^{-} &\rightarrow t \bar{t} \rightarrow b \bar{b} \ell^{-} \ell^{+} \nu_{\ell} \bar{\nu}_{\ell} \, , \\
e^{+} e^{-} &\rightarrow e^{ \pm} \nu_e W^{ \pm} Z \rightarrow e^{ \pm} \nu_e \ell^{\mp} \nu_e b \bar{b} \, , \\ e^{+} e^{-} &\rightarrow \nu_e \bar{\nu}_e Z Z \rightarrow \nu_e \bar{\nu}_e \ell^{-} \ell^{+} b \bar{b} \, , \\
e^{+} e^{-} &\rightarrow Z h, Z \rightarrow \ell^{-} \ell^{+}, h \rightarrow b \bar{b} \, , \\
e^{+} e^{-} &\rightarrow Z W^{+} W^{-}, Z \rightarrow b \bar{b}, W^{+} \rightarrow \ell^{+} \nu_{\ell}, W^{-} \rightarrow \ell^{-} \bar{\nu}_{\ell} \, , \\
e^{+} e^{-} &\rightarrow Z Z Z, Z \rightarrow b \bar{b}, Z \rightarrow \ell^{-} \ell^{+}, Z \rightarrow \nu_{\ell} \bar{\nu}_{\ell} .
\end{align}

Signal and background events are generated using \texttt{MadGraph5\_aMC@NLO}~\cite{Alwall:2014hca} at leading order accuracy in $\alpha_{s}$, interfaced with \texttt{Pythia 8}~\cite{Bierlich:2022pfr} for parton showering, hadronization and multiple parton interactions. The detector efficiencies are simulated using \texttt{Delphes}~\cite{deFavereau:2013fsa} with the CLIC card~\cite{Leogrande:2019qbe}. The jet reconstruction is performed using the Valencia Linear Collider (VLC)~\cite{Boronat:2014hva,Boronat:2016tgd} algorithm with a cone size parameter $R = 1$ at both $\sqrt{s} = 1.5$ and $3$ TeV for the two b-jets. We require a loose b-tagging efficiency of 90$\%$. The events are generated for different charged Higgs masses between 0 GeV and 400 GeV with 10 GeV steps for the signal.

\begin{figure*}[t!]
\centering
 \resizebox{0.65\linewidth}{!}{ \includegraphics{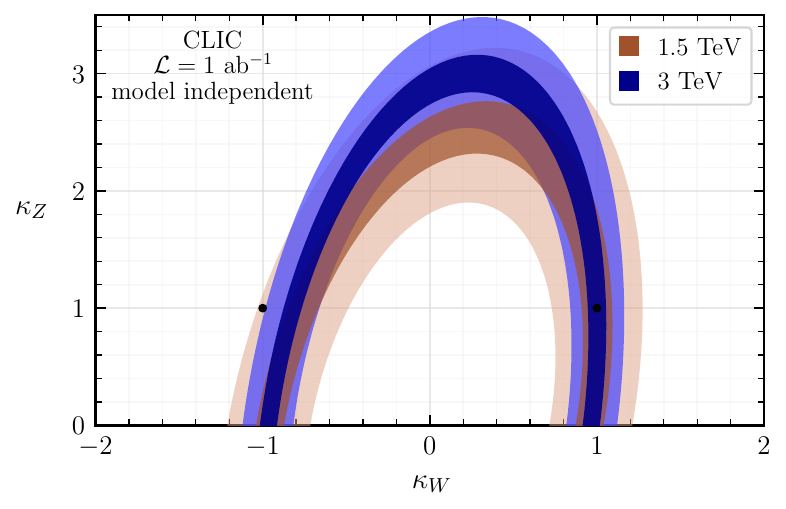}} 
\caption{Model independent sensitivity at CLIC with a luminosity of $\mathcal{L}=1 \text{ab}^{-1}$ and $\sqrt{s} =$ 1.5 (brown) and 3 TeV (blue).  The darker (lighter) colors represent the $2\sigma$ ($5 \sigma$) reach. The value of the charged Higgs mass is chosen to generate the weakest reach for the negative coupling scenario. Excluding this mass translates to an automatic model-independent exclusion of all possible negative coupling models. For this energy scale and polarization, the critical mass is around $300~\text{GeV}$ for both energies. }
\label{fig:moneyplot2}
\end{figure*}
We apply simple kinematic cuts, as the kinematics of the signal and background at these energies are sufficiently different. Some example kinematic distributions can be seen in Figure~\ref{fig:kins}. Interestingly, the kinematics for different mass hypothesis is similar to the SM. This feature can be seen in the bottom of Figure~\ref{fig:kins}, where the difference between the different mass hypothesis and the SM distribution is shown. Because of this, the SM cuts from~\cite{Stolarski:2020qim} are still efficient for this analysis:

\begin{itemize}
\setlength\itemsep{0.0em}
\item $p_{T}^{\ell} > 20~\text{GeV}$.
\item $p_{T}^{j} > 20~\text{GeV}$.
\item one pair of opposite-sign same-flavor leptons
\item Two jets tagged as $b$-jet
\item The invariant mass of the two $b$-jets: $105~\text{GeV} \leq m_{bb} \leq 145 ~\text{GeV}$.
\item The invariant mass of the two isolated leptons  $75~\text{GeV} \leq m_{\ell\ell} \leq 100~\text{GeV}$.
\item The scalar sum of transverse momenta of all reconstructed objects:
\begin{align}
\begin{cases} H_{T} \leq 700~\text{GeV} & \sqrt{s} = 1.5~\mathrm{TeV} \\ H_{T} \leq 1000 ~\text{GeV} & \sqrt{s}  =3~\mathrm{TeV}\end{cases}
\end{align}
\item The missing total transverse momentum: $\slashed{p}_{T} > 10~\text{GeV}$.
\end{itemize}

We assume that the selection efficiency does not change for different values of $\kappa_{W,Z}$ and $m_{H^+}$ as there is no cut depending on it. The number of signal events for different values of model parameters is 
\begin{align}
\mathcal{N}_{S}(\kappa_{W},\kappa_{Z},m_{H^{+}}) = \mathcal{L} \times  \sigma(\kappa_W,\kappa_Z,m_{H^{+}}) \, .
\end{align}

To determine CLIC sensitivity to modified values of gauge-Higgs couplings, we perform a signal rate analysis by constructing a log-likelihood ratio test, assuming that the signal and background event rates follow a Poisson distribution.

The CLIC sensitivity is shown in Fig.~\ref{fig:moneyplot}  in the $\kappa_Z-\kappa_W$ plane for $\mhpl =$ (0, 100, 300) GeV (left, center, right). The brown and blue regions are for $\sqrt{s} = 1.5$ TeV and $\sqrt{s} = 3$ TeV, respectively. The lighter (darker) shaded regions represent the 2$\sigma$ (5$\sigma$) sensitivity. We see that CLIC will exclude the $\lambda_{WZ} = -1$ hypothesis for all values of $m_{H^+}$ with high confidence, at both $\sqrt{s}$ = 1.5 and 3 TeV. 

In Fig.~\ref{fig:moneyplot2}, we show the model-independent sensitivity of CLIC to the negative gauge-Higgs coupling hypothesis. This is obtained by selecting the charged Higgs mass that generates the smallest number of signal events, i.e., with the smallest cross section (see Fig.~\ref{fig:crossDEC}). We again observe that the $\lambda_{WZ} = -1$ hypothesis is excluded with high confidence, implying that any other charged Higgs mass is excluded.

\section{Conclusion} \label{sec:conc}

Understanding the fundamental properties of electroweak symmetry breaking relies on understanding the properties of the Higgs particle. We are reaching the precision era in Higgs physics, but interestingly, we still do not have an experimental determination of the gauge-Higgs coupling sign. These couplings are fundamental in the electroweak sector. Notably, while the ratio of the Higgs couplings to gauge bosons $\lambda_{WZ}$ has been measured at the LHC, the sign has yet to be determined. Measuring the sign requires studying processes with quantum interference, which are also strongly sensitive to new physics.

This paper focuses on $Zh$ production via vector boson fusion (VBF). This process is extremely sensitive to the relative sign of the gauge-Higgs couplings. In the negative coupling scenario $\lambda_{WZ} = -1$, the amplitude for this process exhibits growth with energy and violates unitarity. Taming the high-energy behavior requires the existence of light charged Higgs bosons which can avoid other experimental searches, and thus is virtually hidden from experiments. We probe this elusive light charged scalar using the VBF $Zh$ process and study its effect on the signal event rate and final state kinematics.

We found that a future high-energy lepton collider like CLIC has excellent sensitivity to the negative coupling hypothesis in the presence of light charged Higgs states. We extend previous analyses of this scenario to $m_{H^+} < 370$ GeV, a regime where LHC loses sensitivity ~\cite{deLima:2024uwc}. In particular, $\lambda_{WZ} = -1$ is excluded with high confidence at CLIC with only 1 ab$^{-1}$ of data, independent of the charged Higgs mass. In a different view, if strong deviation from the SM appears in the measurement of this process we can obtain important information on the properties of this charged Higgs state and its role in electroweak symmetry breaking. 

Even though we performed the analysis using CLIC, the same result can be obtained in any future lepton collider with a center-of-mass energy above 1 TeV. A high-energy lepton collider offers a unique opportunity to test the Standard Model at a level never before reached making significant advances in understanding the Higgs sector and the nature of electroweak symmetry breaking.

\section*{Acknowledgments}
We thank David Morrissey, David McKeen, Heather Logan and Daniel Stolarski for their helpful conversations.
This work was supported by the Natural Sciences and Engineering Research Council of Canada (NSERC). C.H.L and D.T is also supported by TRIUMF which receives federal funding via a contribution agreement with the National Research Council (NRC) of Canada.

\appendix

\section{The Effect of Two Charged Higgs Bosons }\label{sec:ap}

It could be possible that the scenario with multiple charged states can give a different prediction from the single state hypothesis explored in this work. In this appendix we show that any bound from multi-state scenarios is always stronger than the model-independent constraint. To show this we consider unitarization in the presence of two charged Higgs states below 370 GeV, with a heavier charged Higgs $H^+_H$ with mass fixed to $m_{H^+_H} = 300$ GeV. We vary the mass $m_{H^+_L}$ of the lighter charged Higgs $H^+_L$.  The sum rule with two states is given by
\begin{align}
g_{WZ}^{L}g_{Wh}^{L} + g_{WZ}^{H}g_{Wh}^{H} = \kappa_{Z}-\kappa_{W} \, .
\end{align}
We define the fractional contribution to unitarization $x_{L,H}$ of the light and heavy states as:
\begin{align}
x_L (\kappa_{Z}-\kappa_{W}) &= g_{WZ}^{L}g_{Wh}^{L} \, , \\
x_H (\kappa_{Z}-\kappa_{W}) &= g_{WZ}^{H}g_{Wh}^{H} \, ,
\end{align}
with $x_L + x_H = 1$. It is sufficient to explore the case $(\kappa_{W},\kappa_{Z})=(1,-1)$ and the effect of varying $x_H$.

\begin{figure}[h]
\centering~\includegraphics[width=1\linewidth]{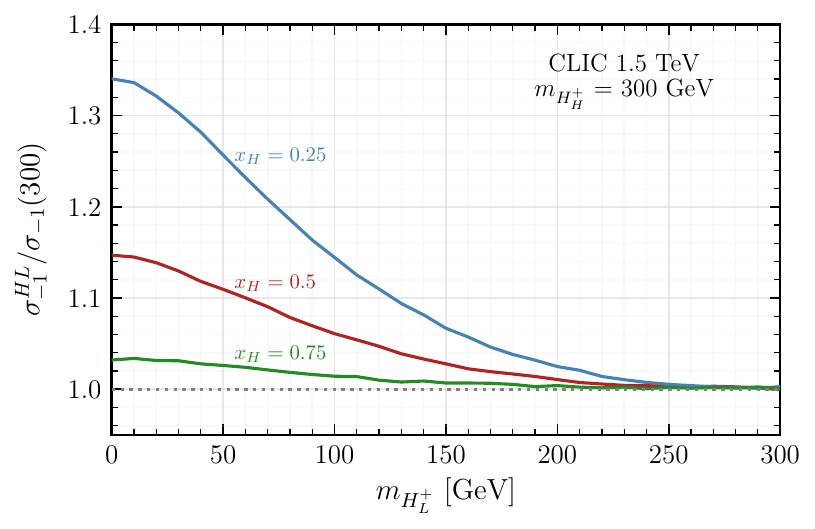} 
\caption{Two-state cross section normalized to the one-state cross section for different values of $x_H$ as a function of the light charged Higgs mass $m_{H^+_L}$. We fix $\kappa_W = -\kappa_Z = 1$ and the heavy charged Higgs mass to $m_{H^+_H} = 300$ GeV.}
\label{fig:twostate}
\end{figure}

In Fig.~\ref{fig:twostate} we show the two-state VBF $Zh$ cross section $\sigma^{HL}_{-1}$ as a function of the light charged Higgs mass $m_{H^+_L}$, normalized to the one-state cross-section, denoted by $\sigma_{-1}$. We see that decreasing $x_H$ increases the cross section for the two-state scenario relative to the one-state scenario, and yields a higher signal rate.  The kinematics do not change substantially as there is no strong interference effect between the two states and thus any bound for this scenario is always stronger than the model-independent bound derived from the one state analysis.  This justifies the model-independence of the one state analysis. 

\bibliographystyle{utphys}
\bibliography{bibZhLC}

\providecommand{\href}[2]{#2}\begingroup\raggedright\begin{thebibliography}{10}

\bibitem{higgsDISCATLAS}
{\bf ATLAS} Collaboration, G.~Aad {\em et al.}, ``{Observation of a new
  particle in the search for the Standard Model Higgs boson with the ATLAS
  detector at the LHC},''
  \href{http://dx.doi.org/10.1016/j.physletb.2012.08.020}{{\em Phys. Lett. B}
  {\bf 716} (2012)  1--29}, \href{http://arxiv.org/abs/1207.7214}{{\tt
  arXiv:1207.7214 [hep-ex]}}.

\bibitem{higgsDISCCMS}
{\bf CMS} Collaboration, S.~Chatrchyan {\em et al.}, ``{Observation of a New
  Boson at a Mass of 125 GeV with the CMS Experiment at the LHC},''
  \href{http://dx.doi.org/10.1016/j.physletb.2012.08.021}{{\em Phys. Lett. B}
  {\bf 716} (2012)  30--61}, \href{http://arxiv.org/abs/1207.7235}{{\tt
  arXiv:1207.7235 [hep-ex]}}.

\bibitem{ATLAS:2022vkf}
{\bf ATLAS} Collaboration, G.~Aad {\em et al.}, ``{A detailed map of Higgs
  boson interactions by the ATLAS experiment ten years after the discovery},''
  \href{http://dx.doi.org/10.1038/s41586-022-04893-w}{{\em Nature} {\bf 607}
  (2022) no.~7917, 52--59}, \href{http://arxiv.org/abs/2207.00092}{{\tt
  arXiv:2207.00092 [hep-ex]}}. [Erratum: Nature 612, E24 (2022)].

\bibitem{CMS:2022dwd}
{\bf CMS} Collaboration, A.~Tumasyan {\em et al.}, ``{A portrait of the Higgs
  boson by the CMS experiment ten years after the discovery.},''
  \href{http://dx.doi.org/10.1038/s41586-022-04892-x}{{\em Nature} {\bf 607}
  (2022) no.~7917, 60--68}, \href{http://arxiv.org/abs/2207.00043}{{\tt
  arXiv:2207.00043 [hep-ex]}}. [Erratum: Nature 623, (2023)].

\bibitem{Chen:2016ofc}
Y.~Chen, J.~Lykken, M.~Spiropulu, D.~Stolarski, and R.~Vega-Morales, ``{Golden
  Probe of Electroweak Symmetry Breaking},''
  \href{http://dx.doi.org/10.1103/PhysRevLett.117.241801}{{\em Phys. Rev.
  Lett.} {\bf 117} (2016) no.~24, 241801},
  \href{http://arxiv.org/abs/1608.02159}{{\tt arXiv:1608.02159 [hep-ph]}}.

\bibitem{Chiang:2018fqf}
C.-W. Chiang, X.-G. He, and G.~Li, ``{Measuring the ratio of HWW and HZZ
  couplings through WWH production},''
  \href{http://dx.doi.org/10.1007/JHEP08(2018)126}{{\em JHEP} {\bf 08} (2018)
  126}, \href{http://arxiv.org/abs/1805.01689}{{\tt arXiv:1805.01689
  [hep-ph]}}.

\bibitem{Stolarski:2020qim}
D.~Stolarski and Y.~Wu, ``{Tree-level interference in vector boson fusion
  production of Vh},''
  \href{http://dx.doi.org/10.1103/PhysRevD.102.033006}{{\em Phys. Rev. D} {\bf
  102} (2020) no.~3, 033006}, \href{http://arxiv.org/abs/2006.09374}{{\tt
  arXiv:2006.09374 [hep-ph]}}.

\bibitem{Xie:2021xtl}
K.-P. Xie and B.~Yan, ``{Probing the electroweak symmetry breaking with Higgs
  production at the LHC},''
  \href{http://dx.doi.org/10.1016/j.physletb.2021.136515}{{\em Phys. Lett. B}
  {\bf 820} (2021)  136515}, \href{http://arxiv.org/abs/2104.12689}{{\tt
  arXiv:2104.12689 [hep-ph]}}.

\bibitem{LHCHiggsXSWG:2013rie}
{\bf LHC Higgs Cross Section Working Group} Collaboration, J.~R. Andersen {\em
  et al.}, ``{Handbook of LHC Higgs Cross Sections: 3. Higgs Properties},''
  \href{http://arxiv.org/abs/1307.1347}{{\tt arXiv:1307.1347 [hep-ph]}}.

\bibitem{Das:2024xre}
D.~Das, A.~Kundu, M.~Levy, A.~M. Prasad, I.~Saha, and A.~Sarkar, ``{Sign of the
  $hZZ$ coupling and implication for new physics},''
  \href{http://arxiv.org/abs/2402.09352}{{\tt arXiv:2402.09352 [hep-ph]}}.

\bibitem{deLima:2024uwc}
C.~H. de~Lima and D.~Stolarski, ``{Influence of new states in searches for
  negative gauge-Higgs couplings},''
  \href{http://arxiv.org/abs/2404.10815}{{\tt arXiv:2404.10815 [hep-ph]}}.

\bibitem{deLima:2021llm}
C.~H. de~Lima, D.~Stolarski, and Y.~Wu, ``{Status of negative coupling
  modifiers for extended Higgs sectors},''
  \href{http://dx.doi.org/10.1103/PhysRevD.105.035019}{{\em Phys. Rev. D} {\bf
  105} (2022) no.~3, 035019}, \href{http://arxiv.org/abs/2111.02533}{{\tt
  arXiv:2111.02533 [hep-ph]}}. [Erratum: Phys.Rev.D 108, 099901 (2023)].

\bibitem{ATLAS:2024vxc}
{\bf ATLAS} Collaboration, G.~Aad {\em et al.}, ``{Determination of the
  relative sign of the Higgs boson couplings to $W$ and $Z$ bosons using $WH$
  production via vector-boson fusion with the ATLAS detector},''
  \href{http://arxiv.org/abs/2402.00426}{{\tt arXiv:2402.00426 [hep-ex]}}.

\bibitem{CMS:2023sdc}
{\bf CMS} Collaboration, \text{CMS Collaboration}, ``{Search for anomalous
  Higgs boson couplings in WH$\to\ell\nu\mathrm{b\overline{b}}$ production
  through Vector Boson Scattering},''.
  \url{https://cds.cern.ch/record/2882655}.

\bibitem{ATLAS:2019mfr}
\text{ATLAS and CMS Collaboration}, ``{Report on the physics at the HL-LHC, and
  perspectives for the HE-LHC: Collection of notes from ATLAS and CMS},''
  \href{http://dx.doi.org/10.23731/CYRM-2019-007.Addendum}{{\em CERN Yellow
  Rep. Monogr.} {\bf 7} (2019)  Addendum},
  \href{http://arxiv.org/abs/1902.10229}{{\tt arXiv:1902.10229 [hep-ex]}}.

\bibitem{Aicheler:2018arh}
{\bf CLIC accelerator} Collaboration, ``{The Compact Linear Collider (CLIC) -
  Project Implementation Plan},'' \href{http://arxiv.org/abs/1903.08655}{{\tt
  arXiv:1903.08655 [physics.acc-ph]}}.

\bibitem{Zarnecki:2020ics}
{\bf CLICdp, ILD concept group} Collaboration, A.~F. Zarnecki, ``{On the
  physics potential of ILC and CLIC},''
  \href{http://dx.doi.org/10.22323/1.376.0037}{{\em PoS} {\bf CORFU2019} (2020)
   037}, \href{http://arxiv.org/abs/2004.14628}{{\tt arXiv:2004.14628
  [hep-ph]}}.

\bibitem{Bhat:2022hdi}
P.~C. Bhat {\em et al.}, ``{Future Collider Options for the US},'' in {\em
  {Snowmass 2021}}.
\newblock 3, 2022.
\newblock \href{http://arxiv.org/abs/2203.08088}{{\tt arXiv:2203.08088
  [hep-ex]}}.

\bibitem{Narain:2022qud}
M.~Narain {\em et al.}, ``{The Future of US Particle Physics - The Snowmass
  2021 Energy Frontier Report},'' \href{http://arxiv.org/abs/2211.11084}{{\tt
  arXiv:2211.11084 [hep-ex]}}.

\bibitem{Bambade:2019fyw}
P.~Bambade {\em et al.}, ``{The International Linear Collider: A Global
  Project},'' \href{http://arxiv.org/abs/1903.01629}{{\tt arXiv:1903.01629
  [hep-ex]}}.

\bibitem{Behnke:2013xla}
``{The International Linear Collider Technical Design Report - Volume 1:
  Executive Summary},'' \href{http://arxiv.org/abs/1306.6327}{{\tt
  arXiv:1306.6327 [physics.acc-ph]}}.

\bibitem{Adolphsen:2013jya}
``{The International Linear Collider Technical Design Report - Volume 3.I:
  Accelerator \textbackslash{}\& in the Technical Design Phase},''
  \href{http://arxiv.org/abs/1306.6353}{{\tt arXiv:1306.6353
  [physics.acc-ph]}}.

\bibitem{Adolphsen:2013kya}
``{The International Linear Collider Technical Design Report - Volume 3.II:
  Accelerator Baseline Design},'' \href{http://arxiv.org/abs/1306.6328}{{\tt
  arXiv:1306.6328 [physics.acc-ph]}}.

\bibitem{Vernieri:2022fae}
C.~Vernieri {\em et al.}, ``{Strategy for Understanding the Higgs Physics: The
  Cool Copper Collider},''
  \href{http://dx.doi.org/10.1088/1748-0221/18/07/P07053}{{\em JINST} {\bf 18}
  (2023) no.~07, P07053}, \href{http://arxiv.org/abs/2203.07646}{{\tt
  arXiv:2203.07646 [hep-ex]}}.

\bibitem{Breidenbach:2023nxd}
M.~Breidenbach, B.~Bullard, E.~A. Nanni, D.~Ntounis, and C.~Vernieri,
  ``{Sustainability Strategy for the Cool Copper Collider},''
  \href{http://dx.doi.org/10.1103/PRXEnergy.2.047001}{{\em PRX Energy} {\bf 2}
  (2023) no.~4, 047001}, \href{http://arxiv.org/abs/2307.04084}{{\tt
  arXiv:2307.04084 [hep-ex]}}.

\bibitem{Andreetto:2024rra}
P.~Andreetto {\em et al.}, ``{Higgs Physics at a $\sqrt{s}=3$ TeV Muon Collider
  with detailed detector simulation},''
  \href{http://arxiv.org/abs/2405.19314}{{\tt arXiv:2405.19314 [hep-ex]}}.

\bibitem{deLima:2024hnk}
C.~H. de~Lima and H.~E. Logan, ``{Can CP be conserved in the two-Higgs-doublet
  model?},'' \href{http://dx.doi.org/10.1103/PhysRevD.110.095007}{{\em Phys.
  Rev. D} {\bf 110} (2024) no.~9, 095007},
  \href{http://arxiv.org/abs/2403.17052}{{\tt arXiv:2403.17052 [hep-ph]}}.

\bibitem{deLima:2024lfc}
C.~H. de~Lima and H.~E. Logan, ``{Is the Real Two-Higgs-Doublet Model
  Consistent?},'' \href{http://dx.doi.org/10.1103/PhysRevLett.133.201801}{{\em
  Phys. Rev. Lett.} {\bf 133} (2024) no.~20, 201801},
  \href{http://arxiv.org/abs/2409.10603}{{\tt arXiv:2409.10603 [hep-ph]}}.

\bibitem{Alwall:2014hca}
J.~Alwall, R.~Frederix, S.~Frixione, V.~Hirschi, F.~Maltoni, O.~Mattelaer,
  H.~S. Shao, T.~Stelzer, P.~Torrielli, and M.~Zaro, ``{The automated
  computation of tree-level and next-to-leading order differential cross
  sections, and their matching to parton shower simulations},''
  \href{http://dx.doi.org/10.1007/JHEP07(2014)079}{{\em JHEP} {\bf 07} (2014)
  079}, \href{http://arxiv.org/abs/1405.0301}{{\tt arXiv:1405.0301 [hep-ph]}}.

\bibitem{Bierlich:2022pfr}
C.~Bierlich {\em et al.}, ``{A comprehensive guide to the physics and usage of
  PYTHIA 8.3},'' \href{http://dx.doi.org/10.21468/SciPostPhysCodeb.8}{{\em
  SciPost Phys. Codeb.} {\bf 2022} (2022)  8},
  \href{http://arxiv.org/abs/2203.11601}{{\tt arXiv:2203.11601 [hep-ph]}}.

\bibitem{deFavereau:2013fsa}
{\bf DELPHES 3} Collaboration, J.~de~Favereau, C.~Delaere, P.~Demin,
  A.~Giammanco, V.~Lema\^\i{}tre, A.~Mertens, and M.~Selvaggi, ``{DELPHES 3, A
  modular framework for fast simulation of a generic collider experiment},''
  \href{http://dx.doi.org/10.1007/JHEP02(2014)057}{{\em JHEP} {\bf 02} (2014)
  057}, \href{http://arxiv.org/abs/1307.6346}{{\tt arXiv:1307.6346 [hep-ex]}}.

\bibitem{Leogrande:2019qbe}
E.~Leogrande, P.~Roloff, U.~Schnoor, and M.~Weber, ``{A DELPHES card for the
  CLIC detector},'' \href{http://arxiv.org/abs/1909.12728}{{\tt
  arXiv:1909.12728 [hep-ex]}}.

\bibitem{Boronat:2014hva}
M.~Boronat, J.~Fuster, I.~Garcia, E.~Ros, and M.~Vos, ``{A robust jet
  reconstruction algorithm for high-energy lepton colliders},''
  \href{http://dx.doi.org/10.1016/j.physletb.2015.08.055}{{\em Phys. Lett. B}
  {\bf 750} (2015)  95--99}, \href{http://arxiv.org/abs/1404.4294}{{\tt
  arXiv:1404.4294 [hep-ex]}}.

\bibitem{Boronat:2016tgd}
M.~Boronat, J.~Fuster, I.~Garcia, P.~Roloff, R.~Simoniello, and M.~Vos, ``{Jet
  reconstruction at high-energy electron\textendash{}positron colliders},''
  \href{http://dx.doi.org/10.1140/epjc/s10052-018-5594-6}{{\em Eur. Phys. J. C}
  {\bf 78} (2018) no.~2, 144}, \href{http://arxiv.org/abs/1607.05039}{{\tt
  arXiv:1607.05039 [hep-ex]}}.

\end{thebibliography}\endgroup

\end{document}